**В. В. Писляков**

*Национальный исследовательский университет «Высшая школа экономики»,*

*Москва, Россия*

# Самоцитирование и его влияние на оценку научной деятельности: обзор литературы. Часть I

**Аннотация:** В статье представлен обзор литературы, посвященной влиянию самоцитирования и возникающим от этого возможным искажениям при библиометрическом анализе. Вводится обобщенное определение самоцитирования и его частных вариантов: авторского, институционального, странового, журнального, дисциплинарного, издательского. Приводятся формулы основных метрик самоцитирования — коэффициентов самоцитирования и самоцитируемости. Далее подробно рассматривается мировая литература по авторскому, институциональному, страновому и журнальному самоцитированию. Обобщаются текущие взгляды на роль и влияние самоцитирования при оценке научной деятельности. При аналитическом рассмотрении статей, посвященных самоцитированию, выясняется, что у исследователей существует консенсус по ряду позиций, например: (а) патологией является как гипертрофированное самоцитирование, так и его отсутствие; (б) самоцитирование мало влияет на оценку крупных научных единиц, но может быть критическим при анализе отдельных ученых; (в) влияние самоцитирования наиболее выражено у научных единиц со слабыми библиометрическими показателями, в то время как топовые ученые, организации, журналы и др. получают наибольшее число ссылок извне. В обзоре также рассмотрено реагирование самих библиометрических инструментов и баз данных с целью корректировки индикаторов в случае манипулирования самоцитированием.

Представленная здесь первая часть обзора посвящена определению основных понятий и терминов, а также рассмотрению наиболее обсуждаемого и распространенного типа самоцитирования — авторского.

**Ключевые слова:** самоцитирование, библиометрический анализ, оценка научной деятельности, импакт-фактор.

**Vladimir Pislyakov**

*HSE University, Moscow, Russia*

**Self-citation and its impact on research evaluation:**

**Literature review. Part I**

**Abstract:** This review summarizes papers which analyze impact of self-citation on research evaluation. We introduce a generalized definition of self-citation and its variants: author, institutional, country, journal, discipline, publisher self-citation. Formulae of the basic self-citation measures are given, namely self-citing and self-cited rates. World literature on author, institutional, country and journal self-citation is studied in more detail. Current views on the role and impact of self-citation are compiled and analyzed. It is found that there is a general consensus on some points: (a) pathological is as excessive self-citation so its total absence; (b) self-citation has low impact on large science units but may be critical for analysis of individual researchers; (c) share of self-citations is generally higher for units with low bibliometric performance, while top scientists, institutions, journals receive the majority of their citations from outside. This review also considers how bibliometric instruments and databases respond to challenge of possible manipulation by self-citations and how they correct bibliometric indicators calculated by them.

The first part of the review presented here deals with the fundamental terms and definitions, and the most discussed and studied type of the self-citation, author self-citation.

The paper was funded by RFBR, project number 20-111-50209.

Оценка результативности и эффективности научной деятельности при помощи библиометрических методов в настоящее время получила широкое распространение по всему миру. Ввиду того, что нередко такая оценка прямо или косвенно влияет на карьеры ученых, распределение финансирования, внесение изменений в структуру научных и образовательных организаций, повышенное внимание уделяется тщательности и корректности наукометрического анализа, отсутствию эмпирических или методологических искажений. Одним из наиболее известных, часто упоминаемых искажающих факторов называется самоцитирование. Ссылками на собственные научные работы можно «раздуть» библиометрические показатели автора, подразделения, организации и т. д.

Опасения о возможности злоупотребления самоцитированием высказывались практически с самого старта эры индексирования библиографических ссылок и появления Science Citation Index [**1–3**], также в некоторых исследованиях сразу использовались методологические поправки на самоцитирование [**4**].

Данная статья содержит обзор подходов к исследованию и корректировке библиометрического самоцитирования. При этом мы воздержимся от рассмотрения этических вопросов, «позволительности» того или иного библиографического поведения, нас будет интересовать только научная сторона и выявление грамотных стратегий аналитика при работе с самоцитированием. Насколько известно автору, это первый обзор как в отечественной, так и в мировой литературе, посвященный выделенной теме самоцитирования в науке. Из существующих трудов можно отметить небольшие разделы, посвященные самоцитированию, в общих обзорах литературы по информетрии Д. Бар-Илан [**5**. P. 14–15] и индикаторам цитируемости Л. Уолтмана [**6**. P. 373–374], исторический экскурс в [**7**. P. 220–222, 260], конспект ряда статей по данной теме в [**8**], а также ёмкое введение с обширным списком литературы в [**9**].

На текущий момент (июнь 2021 г.) в базе данных Scopus содержатся 259 документов, только *в названии*, заголовке которых есть словосочетание «self-citation(s)». В Web of Science Core Collection таких источников 261[1]. Очевидно, что работ, так или иначе затрагивающих самоцитирование, в разы больше. Поэтому настоящий обзор не претендует на исчерпывающий охват, в нем собраны журнальные статьи, отражающие разные взгляды и эмпирические наблюдения за самоцитированием, ярко подсвечивающие многообразие данной проблематики и различные методологические подходы.

Практически все источники, вошедшие в обзор, мы снабдили Интернет-ссылками в списке литературы, что даст возможность читателю оперативно ознакомиться с рассматриваемыми здесь работами. Часть этих трудов находится в открытом доступе, часть доступна по подписке, которая в настоящее время организована во многих российских научных и образовательных организациях, в том числе в рамках централизованного/национального доступа по программе Минобрнауки РФ.

## Определение и типы самоцитирования, основные понятия

Для строгости изложения начнем наш обзор с наиболее общего определения библиометрического самоцитирования (*self-citation*, реже *self-reference*). Оно может быть сформулировано так: самоцитирование — это библиографическая ссылка в одной публикации на другую публикацию, если у этих публикаций есть некий объединяющий признак. Чаще всего это означает, что в создании как первой, так и второй научной работы принимала

---

участие одна и та же «исследовательская единица» или обе опубликованы в одном журнале/одной дисциплинарной области. Соответственно, мы будем разделять следующие типы самоцитирования:

- *авторское самоцитирование* — цитирующий и цитируемый документы созданы одним и тем же (со)автором [**10–11**];

- *институциональное самоцитирование* — обе публикации написаны сотрудниками одной и той же лаборатории, исследовательской группы, института, университета [**12–14**];

- *страновое самоцитирование* — в создании источника и адресата ссылки принимали участие ученые одной и той же страны (точнее: работающие в одной стране) [**15–16**];

- *журнальное самоцитирование* — обе работы опубликованы в одном журнале [**17.** P. 124];

- *дисциплинарное самоцитирование* — цитирующая и цитируемая публикации относятся к одной и той же области науки (по той или иной принятой в рамках конкретного библиометрического исследования классификации) [**18–19**];

- *издательское самоцитирование* — ссылка из статьи журнала, выпускаемого в некотором издательстве, ведет на статью журнала, который публикуется тем же издательством [**20**].

Подобное разделение проводят, например, в [**11**; **7.** P. 220] (за исключением пунктов «страновое» и «издательское» самоцитирование). Кроме этого, иногда выделяют «языковое самоцитирование» [**21–22**], однако оно, дисциплинарное и издательское исследуются редко, поэтому мы не будем их затрагивать в нашем обзоре.

Обобщая: *если задано условие, по которому создаются подмножества документов, удовлетворяющих этому условию, то библиографическая ссылка, ведущая из документа, принадлежащего одному из таких подмножеств, на* *документ, также принадлежащий тому же подмножеству, является самоцитированием по заданному условию*.

Заметим, что необходимо отличать анализ, например, странового самоцитирования (исследуются ссылки ученых страны на ученых той же страны) от анализа влияния авторского самоцитирования на страновые (национальные) показатели: как ссылки авторов на самих себя повышают уровень цитируемости всей страны. В последнем случае речь идет все-таки именно об авторском самоцитировании.

Далее приведем два показателя, которые будут встречаться и использоваться на протяжении всего данного обзора. Вслед за [**23–24**] сначала определим их для самого простого контекста, для журнального самоцитирования. В числителе обоих индикаторов — число ссылок, полученных журналом из статей, опубликованных в нем самом за определенный период исследования. Число «самоцитирований». Знаменатели двух индикаторов отличаются: в знаменателе первого — число всех ссылок, *полученных* журналом за тот же период (заметьте: включая и самоцитирования). Этот показатель называется коэффициентом самоцитируемости, он показывает долю во всех ссылках, полученных журналом, ссылок, которые получены им из себя самого. Знаменатель второго показателя равен числу всех ссылок, *сделанных* журналом (включая самоцитирования). Он называется коэффициентом самоцитирования и показывает долю во всех цитированиях, сделанных журналом, ссылок, которые ведут на него самого. Графически это проиллюстрировано на рис. 1.

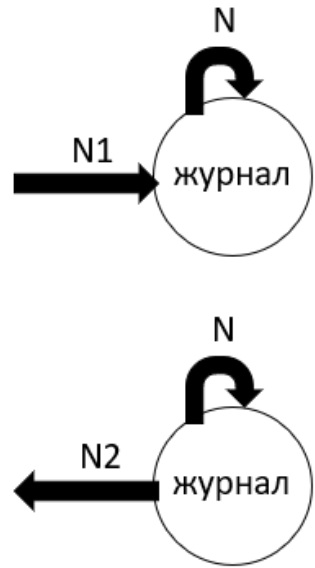

**Рис. 1**. Коэффициенты самоцитируемости и самоцитирования. N — число цитирований журналом самого себя (самоцитирований). Вверху: N1 — число «входящих» ссылок извне (из других изданий). Коэффициент самоцитируемости = N / (N+N1). Внизу: N2 — число «исходящих» ссылок вовне (на другие издания). Коэффициент самоцитирования = N / (N+N2).

В случае авторского самоцитирования ситуация несколько более сложная. Для его оценки применяются аналогичные коэффициенты, в первую очередь коэффициент самоцитируемости, однако в этом случае есть два различных подхода к тому, что именно считать самоцитированием. «Прямое самоцитирование» — это ссылка на публикацию, среди соавторов которой есть некоторый исследуемый ученый, приведенная в статье, среди соавторов которой есть тот же самый ученый. Однако есть еще и самоцитирование соавторов — ссылка на публикацию, среди соавторов которой есть исследуемый ученый, приведенная в статье, среди соавторов которой есть другие соавтор(ы) цитируемой статьи (при этом самого исследуемого ученого в авторах цитирующей публикации может не быть). Поясним подробнее [**24. С. 193**]: пусть ученый S написал (сам или в соавторстве) статью P1, которую цитирует статья P2, P2 → P1. Это считается:

• прямым самоцитированием, если S — автор P2;

• самоцитированием соавторов, если среди авторов P2 есть хотя бы один из авторов P1.

Соответственно, коэффициент самоцитируемости ученого может считаться или как доля «прямого самоцитирования» во всех ссылках, которые получили его публикации, или как доля «самоцитирования соавторов» во всех ссылках, полученных его работами. Очевидно, что второй коэффициент будет всегда больше или равен первому.

Вообще говоря, исключение при анализе работ ученого всех ссылок, являющихся «самоцитированием соавторов», предлагает весьма ригористичный подход: учтем только те цитирования, которые получены из «совсем чужих» работ — множество авторов которых никак не пересекается с множеством авторов цитируемой публикации. А всё остальное посчитаем самоцитированием. Поэтому, как отмечают авторы [**25**], при анализе конкретного ученого авторским самоцитированием логичнее считать только «прямое самоцитирование». Однако при переходе на «мезо-» или «макро-уровень», т. е. при анализе факультетов, организаций или целых стран, это определение неприменимо. В результате авторы [**25**], анализирующие страны и весь мировой публикационный поток, следуют логике второго определения, согласно которому самоцитированием считается любой случай, когда множество авторов цитирующей и цитируемой статьи пересекаются хотя бы по одному автору.

Кроме того, была предпринята интересная попытка примирить оба подхода: [**26**] предлагает «фракционный», долевой подсчет самоцитирования, при котором ссылка одной статьи на другую считается самоцитированием с некоторым коэффициентом от 0 (множества авторов цитирующей и цитируемой статьи не пересекаются) до 1 (полностью идентичный состав авторов). В качестве меры «степени самоцитирования» здесь предлагается коэффициент Жаккара.

Авторы фундаментального труда «Introduction to Informetrics» [**7**], классики информетрии Л. Эгге и Р. Руссо, как уже было упомянуто, проводят разделение на разные типы самоцитирования, подобные введенному в нашей работе (за исключением внутристранового и издательского самоцитирования). Они подчеркивают, что учет «институционального самоцитирования» проводится, прежде всего, при использовании библиометрического анализа в контексте научной политики. Однако по поводу авторского самоцитирования они ограничиваются лишь обзором литературы (p. 220–222). Более самостоятельная информация содержится в блоке, посвященном журнальному самоцитированию (p. 260). Здесь дается краткое определение коэффициенту самоцитирования и самоцитируемости журналов, с методикой их определения по Journal Citation Reports, JCR (на тот момент — не базы данных, а ежегодной печатной публикации Института научной информации, ISI). Авторы предлагают любопытную трактовку высокого значения этих индикаторов, свободную от всяких этических суждений: высокий коэффициент самоцитируемости свидетельствует о малой заметности журнала; высокий коэффициент самоцитирования говорит о том, что журнал относится к замкнутой, изолированной научной дисциплине. В статьях [**27–28**] один из авторов с вариациями повторяет эти рассуждения, поясняя, что, как следствие, высокая самоцитируемость скорее характерна для периферийных журналов (а низкая — для лидирующих). Высокий процент самоцитирования свойственен более специализированным изданиям.

**Авторское самоцитирование**

Р. Тальякоццо проводит одно из первых подробных исследований авторского самоцитирования, для чего тщательно выделяет его из других типов самоцитирования — в рамках одного журнала, одной организации и одной области науки [**11**]. Набор данных ограничивается шестью журналами, из которых три по физиологии растений и три по нейробиологии (в сумме 183 статьи). Средние коэффициенты авторского самоцитирования получены 16,6% и 17,5% соответственно. Авторы в среднем цитируют самих себя ощутимо активнее, чем любого другого автора, причем в двух смыслах. И «библиографически» — если судить по списку литературы, в нем чаще встречаются работы авторов статей, чем произведения «внешних» авторов; и по числу внутренних отсылок в тексте на каждую конкретную работу (многократное упоминание одной и той же статьи на протяжении цитирующей публикации). Последнее обстоятельство [**11**] интерпретирует как доказательство более тесной смысловой связи цитируемой статьи с цитирующей при наличии у них общего автора: цитируемая статья упоминается несколько раз по всему тексту цитирующей работы. Также на исследуемом примере обнаружено, что при самоцитировании обычно цитируется более свежая, недавняя литература, чем при цитировании других авторов. В то же время не найдено статистически значимой зависимости объема авторского самоцитирования от числа соавторов, публикационной продуктивности автора и длины списка литературы в статьях.

А. Портер [**29**] подробно исследует публикации и цитируемость 69-ти психологов (данные детально описаны в [**30**]). Он определяет самоцитирование в широком смысле, объединяя прямое самоцитирование и самоцитирование соавторов. Доля таких самоцитирований среди всех 1920 ссылок, вошедших в исследование, составила 16,1%. Прежде всего, автор констатирует, что корреляция между показателями ученых с учетом самоцитирований и без их учета предельно высока, 0,99 (коэффициент Пирсона). Если считать для каждого ученого только статьи, где он первый (или единственный) автор, коэффициент практически не меняется — 0,98. Таким образом, делается вывод, что для некоторых целей («for some purposes») вполне можно не делать поправку на самоцитирование. Однако данный показатель неоднороден: у 10% ученых доля самоцитирований достигает и даже превышает 30%. Психология — комплексная дисциплина, и А. Портер анализирует ее различные направления, приходя к выводу, что чем ближе суб-дисциплина в психологии к естественным наукам, тем ниже уровень самоцитируемости. Он ссылается на А. Медоуза [**31**], который предполагал, что низкий уровень самоцитируемости свидетельствует о «зрелости» дисциплины, и распространяет это утверждение на отдельных ученых: число и доля «не-самоцитирований» может отражать процесс достижения ученым зрелости, когда его работы начинают замечать другие исследователи.

Во влиятельной работе [**32**] (к июню 2021 г. 188 полученных ссылок в Scopus) исследуется роль *авторского* самоцитирования в цитируемости работ, написанных в *институциональном* международном сотрудничестве. Используются данные из мониторинга литературы по астрономии, вышедшей в Голландии за 12 лет [**33**]. Известно, что статьи, написанные в соавторстве, в среднем цитируются больше, чем «соло-статьи». Однако, и здесь А. ван Раан оппонирует автору [**34**], это преимущество цитируемости не может быть объяснено только повышенным самоцитированием таких статей. Действительно, работы, написанные в авторском коллективе, обычно получают больше самоцитирований (потому что авторов больше), однако только этим позитивный библиометрический эффект от научного сотрудничества объяснен быть не может. Заметим, однако, что в [**32**] также содержатся данные о том, что внутреннее (внутри одной страны) институциональное сотрудничество приводит к снижению средней цитируемости статьи по сравнению с работами в рамках одной организации, поэтому выводы данного исследования справедливы именно для международных институциональных коллабораций.

Д. Акснес исследует влияние авторского самоцитирования на национальном уровне, анализируя цитируемость более 45 тыс. публикаций норвежских ученых за 16 лет [**35**]. Выясняется, что обычно наиболее цитируемые работы также получают наибольшее число самоцитирований, но при этом *процент*, *доля* самоцитирований в полученных ими ссылках меньше. Кроме того, чем больше у статьи авторов, тем в среднем выше число самоцитирований, но доля самоцитирований меньше — т. е. более высокая цитируемость работы, написанной большим коллективом, обеспечивается не только и не столько ссылками, полученными от участников этого коллектива (здесь вновь полемика с [**34**]). Средний коэффициент самоцитируемости варьируется по научным дисциплинам от 17 до 31%, причем чем ближе дисциплина к естественным наукам, тем он выше (здесь некоторое несовпадение с Портером [**29**]). Это дает автору повод предположить, что в таких науках выше «кумулятивность», построение ученым нового знания на основании своих предыдущих работ. Наконец, доля самоцитирования высока в полученных статьей ссылках за первый-второй год после ее выхода, затем начинают преобладать «внешние» цитирования. В качестве обобщающего вывода автор говорит об отсутствии критического влияния со стороны самоцитирования при библиометрическом анализе на макроуровне (исследование целых стран). Однако на более низком уровне агрегирования, при оценке исследовательских групп и отдельных ученых, самоцитирование представляет серьезную проблему при формировании научной политики. Для корректного анализа оно должно исключаться из расчетов.

В. Гленцль с соавторами комплексно изучают авторское самоцитирование во всех статьях 1992 г. выхода, проиндексированных базой данных Web of Science [**25**]. Они начинают с утверждения, что в библиометрии общепризнанно считается «патологическим» как полное отсутствие самоцитирования, так и его гипертрофированность. Одна из целей работы — изучить закономерности самоцитирования на больших данных, чтобы приблизиться к представлению о его «стандартной модели», наиболее часто встречающихся свойствах и характеристиках. Тогда при оценке научной деятельности можно будет обращать особое внимание на значительные отклонения от выявленной модели. Авторы обнаруживают: (1) Число самоцитирований во всех дисциплинах растет быстрее, чем число «внешних» ссылок, достигая максимума на следующий год или через два года после публикации, в то время как пик внешних ссылок наблюдается 1–2 года позднее. Однако самоцитирования также гораздо быстрее затухают, их кумулятивный процент уменьшается с ~50% в год выхода статьи до ~20% через 10 лет. (2) Полученные самоцитирования не являются ни строго зависимыми, ни абсолютно независимыми от числа полученных статьёй внешних ссылок. Ближе всего зависимость наиболее вероятного числа самоцитирований у статьи от полученных ею на заданный момент времени внешних ссылок может быть описана степенной функцией с показателем степени ~0,55, т. е. число самоцитирований примерно равно квадратному корню полученных «извне» ссылок. (3) Если объединять данные по странам, то окажется, что наибольший процент авторских самоцитирований в статьях тех стран, чьи авторы чаще публикуются в низкоцитируемых журналах, а показатели цитируемости стран как целого тоже ниже. Авторы [**25**] подводят итог: низкая заметность (visibility) публикаций ведет к тому, что ученые чаще цитируют себя сами, чем получают ссылки от других, вероятность самоцитирования увеличивается.

Методика предыдущего исследования воспроизведена в [**36**], но для более свежих данных (публикации 2000 г.) и на макроуровне — исследуется влияние авторского самоцитирования на показатели государств как целого. Работа сфокусирована на различиях, наблюдаемых на уровне стран при сравнении индикаторов, которые посчитаны до и после исключения из анализа самоцитирования (оно берется в широком смысле — и прямое и соавторское). Для всех естественных/технических/медицинских наук, рассмотренных как единое целое, подтверждается вывод [**25**]. Существует сильная, статистически значимая зависимость: чем в менее цитируемых журналах публикуются в среднем ученые страны, тем выше доля получаемых авторских самоцитирований у страны как целого. Публикации стран, чьи авторы много публикуются в лидирующих журналах, имеют в среднем более низкий коэффициент самоцитируемости. Если построить регрессионную прямую по данным цитируемости стран с учетом и без учета самоцитирований, то будет получен средний коэффициент самоцитируемости, он равен 26%. На взаимное расположение 35 ведущих стран, включенных в исследование, «вычищение» самоцитирований влияет несильно (анализируется показатель relative citation rate — цитирование статей страны относительно среднего уровня журналов, в которых они опубликованы). Далее авторы тщательно исследуют влияние самоцитирования по 15 крупным научным дисциплинам и приходят к выводу, что в целом на макроуровне целых стран исключение из расчетов самоцитирований несильно влияет на выводы библиометрического анализа, а значит для этого рода задач необязательно проводить такую коррекцию. Однако есть и некоторые различия по дисциплинам — например, обратная зависимость доли авторских самоцитирований и среднего уровня журналов, где публикуются авторы страны, особенно чётко прослеживается для биологии и химии. В то же время для ряда медицинских специальностей, а также общественных и гуманитарных наук она практически отсутствует (в случае некоторых социальных областей она даже, наоборот, слабопозитивная). В исследование включена Россия, для которой обнаружены выделяющиеся на общем фоне высокие коэффициенты самоцитируемости в биомедицинских науках, медицине, нейронауках, химии и технических науках. При этом неожиданно низкий коэффициент авторской самоцитируемости наблюдается у отечественных общественно-научных работ. Заметим, что, к сожалению, статья [**36**] содержит немало ошибок, в основном в графическом представлении результатов и обозначении средних уровней, однако текстовая часть в целом корректна и не вызывает сомнений в сделанных авторами выводах.

В своей другой работе, которая посвящена библиометрии более детального «мезо-уровня», Б. Тайс и В. Гленцль исследовали показатели двенадцати европейских университетов разной специализации из шести стран [**37**]. Они пришли к выводу, что изменение их библиометрических показателей при исключении из анализа авторских самоцитирований разнообразно и не повторяет в точности характер изменений, которые происходят при аналогичной корректировке индикаторов как стран, в которых расположены университеты, так и научных дисциплин, по которым они специализируются. В отличие от [**36**], где в целом делается вывод об избыточности коррекции на авторское самоцитирование при макроанализе целых стран, здесь, на мезо-уровне организаций, рекомендуется одновременно учитывать и анализировать как показатели, включающие самоцитирование, так и показатели с исключенными ссылками (со)авторов на свои статьи.

Работа [**38**] изучает вопрос влияния числа соавторов на абсолютное число и процент самоцитирований, которое получает статья (исследуются трех- и десятилетнее окна цитирования). В статье предлагается простая вероятностная математическая модель зависимости доли самоцитирования от размера авторского коллектива, которая оказывается вариантом степенной функции, асимптотически стремящейся к 1 (очевидно: при бесконечном авторском коллективе все ссылки — самоцитирования). Далее эта модель сопоставляется с эмпирическими данными, которые лишь частично оправдывают моделирование: в реальности коэффициент самоцитируемости уже начиная с 10 авторов практически не меняется. При этом абсолютное число полученных самоцитирований растет с ростом числа авторов значительно медленнее, чем число «внешних» ссылок. Авторы [**38**] делают вывод: расхожее мнение о том, что увеличение числа соавторов прежде всего увеличивает самоцитирование, не подтверждается наблюдениями. Соавторство увеличивает, прежде всего, вероятность быть процитированными другими учеными — не входящими в авторский коллектив. Что касается доли полученных ссылок, коэффициент самоцитируемости ощутимо меньше только для статей с одним автором. Для случая статей, написанных вообще без соавторов, утверждение о низкой доле самоцитирований подтверждается.

Ряд недавних статей посвящён изменению поведения учёных *до* и *после* введения государственных мер по оценке публикационной активности/продуктивности, если в системе оценки не делается разницы между самоцитированием и «внешним» цитированием. Чаще всего это демонстрируется на примере Италии с программой National Scientific Accreditation [**39–41**]. Обзор результатов проведен в [**42**].

Из новых подходов интересен оригинальный *s*-индекс, сделанный по образу индекса Хирша [**43**]. Определение идентичное: «*s*-индекс показатель учёного равен *s*, если каждая из *s* его публикаций получила не менее *s* самоцитирований» [**44**]. Конечно, его смысл, в отличие от *h*-индекса, трактуется скорее в негативном ключе. Применение данного индекса ограничено (как и индекса Хирша), но взгляд заслуживает изучения.

Институциональному, страновому и журнальному самоцитированию будет посвящена вторая часть настоящей работы.

**Список источников**

**Информация об авторе**
**Писляков Владимир Владимирович** – канд. физ.-мат. наук, заместитель директора библиотеки НИУ «Высшая школа экономики», член редколлегии «Journal of Informetrics»
ORCID: 0000-0002-4889-9858
pislyakov@hse.ru
**Information about the author**
**Vladimir Pislyakov** – PhD, Assistant Library Director, HSE University; Editorial board member for Journal of Informetrics
ORCID: 0000-0002-4889-9858
pislyakov@hse.ru